\documentclass{iopart}
\usepackage{graphicx}

\begin{document}

\title{Evidence for chemical equilibration at RHIC}
\author{Daniel Magestro}
\address{Gesellschaft f\"ur Schwerionenforschung, D-64291 Darmstadt, Germany}


\begin{abstract}

This contribution focuses on the results of statistical model calculations at
RHIC energies, including recently available experimental data.  Previous
calculations of particle yield ratios showed good agreement with measurements
at SPS and lower energies, suggesting that the composite system possesses a
high degree of chemical equilibrium at freeze-out.  The effect of feeddown
contamination on the model parameters is discussed, and the sensitivity of
individual ratios to the model parameters ($T$, $\mu_B$) is illustrated.

\end{abstract}

\section{Introduction}

Over the last few years, the study of thermalization in heavy-ion collisions
using statistical models has been extended to SPS
\cite{pbm1999plb,becattini2001,kabana2001} and RHIC
\cite{pbm2001plb,krakow2001} energies, allowing the so-called ``freeze-out
curve" \cite{cleymans1999} of hadronic matter to be extended close to the
$\mu_B=0$ axis of the QCD phase diagram.  This is an important development in
the search for the quark gluon plasma because it allows a more direct
comparison between lattice QCD calculations of the phase boundary temperature
\cite{karsch2001}, performed mostly at zero baryon density, and
experiment-based methods for measuring the equilibriun temperature on the
hadronic side of the phase boundary (if the system equilibrates).

The updated phase diagram presented at this conference by Braun-Munzinger
\cite{pbmSQM} suggests a scenario in which the final-state composition of the
system, determined at chemical freeze-out, relies less on hadronic
rescattering processes \cite{bass2000prc} as the beam energy is increased. This
effect arises due to two reasons: stronger transverse flow causes faster
expansion of the system, allowing less time for inelastic interactions; and
the hadronic phase is bounded above by the temperature of the phase boundary
itself, meaning that the system cools very little between hadronization and
freeze-out.

In our previous paper \cite{pbm2001plb}, we showed that particle ratios at
midrapidity measured during the RHIC run at $\sqrt{s} = 130$ GeV are
consistent with a system in chemical equilibrium, as described by a
statistical model. The best-fit model parameters ($T=174 \pm 7$ MeV, $\mu_B=46
\pm 6$ MeV) differ by a factor 5 in baryon chemical potential $\mu_B$ from the
fit at maximum SPS energy \cite{pbm1999plb}, while the temperature changes very
little (SPS: $T=168 \pm 3$ MeV, $\mu_B=266 \pm 5$ MeV). The small change in
temperature, together with the good agreement with lattice QCD predictions of
$T_c$ \cite{karsch2000} ($T_c = 170 - 190$ MeV), suggests that the system is
nearly borne into chemical equilibrium out of the phase transition
\cite{pbm1996npa,stock1999plb}.  We also presented predictions for particle
ratios at full RHIC energy $\sqrt{s} = 200$ GeV.

In this contribution, the model comparison for the RHIC run at $\sqrt{s} =
130$ GeV is extended to include additional ratio measurements which have been
recently published or presented, including at this conference.  The treatment
and effect of feeddown correction to particle densities in the model is
discussed. Finally, the question of which ratios drive the model fit to the
data is investigated.  It is worth noting that the model parameter values
($T$, $\mu_B$) obtained here for the best fit to the data have not changed
from the fit performed with the data available at the time of our publication.

\section{Overview of experimental results}\label{sec:expt}

The primary tool for investigating chemical equilibration in heavy-ion
collisions is the experimental measurement of particle production, and in
particular ratios of particle species. All four RHIC experiments have
published or presented preliminary results on particle ratios.  Some
experimental observations are \cite{2001qm}:

\begin{itemize}
\item +/- ratios are approaching unity at midrapidity, indicating that quark coalescence and pair creation processes
play a greater role in final-state particle yields.  However, the midrapidity
region is not yet baryon free at $\sqrt{s} = 130$.
\item No large trends have been reported in ratios as a function of $p_{t}$ in the relevant $p_{t}$ range.
\item $\overline{p}/p$ increases from $\approx 0.4$ at $y=2$ to $\approx 0.65$ at
midrapidity, with a plateau width of 2 rapidity units. $\pi^{-}/\pi^{+}$ shows
no rapidity dependence over the same region.
\end{itemize}

Since the submission of our paper \cite{pbm2001plb}, several additional
measurements of ratios have been presented for the RHIC 2000 data.  The $\phi
/ h^{-}$ ratio was measured by the STAR Collaboration \cite{STARphiprl} to be
$0.021 \pm 0.001_{\rm stat} \pm 0.004_{\rm sys}$.  The PHENIX Collaboration
presented \cite{morrisonSQM} preliminary results for $K^{-}/\pi^{-}$ at this
conference (most central:  $K^{-}/\pi^{-} = 0.15 \pm 0.01_{\rm stat} \pm
0.03_{\rm sys}$).

Also at this conference, an updated preliminary value for $K^{\star0}/h^{-}$
was presented \cite{fachiniSQM} by the STAR Collaboration ($|K^{\star0}|/h^{-}
= 0.032 \pm 0.003_{\rm stat} \pm 0.008_{\rm sys}$). The new value differs from
the previous value \cite{zxu2001} by nearly a factor 2 due to improved
statistics and a different treatment of the residual background.

\section{Statistical model calculations}

The statistical model employed in the present calculations was first described
in \cite{pbm1999plb}.  In the model, particle densities are calculated as a
grand canonical ensemble with the requirement that baryon, strangeness and
isospin quantum numbers are conserved. All known particles and resonances up to
2 GeV are included in the calculation, and particles decay according to their
branching ratios.

The best fit model parameters ($T$, $\mu_B$) are determined by comparing model
ratios to experimental values and finding the minimum $\chi^2$.  A
$\chi^2/{\rm d.o.f.\sim 1}$ indicates a good fit.  When both statistical and
systematic errors have been presented for experimental values, the errors are
added in quadrature.

In this work, the best fit parameters are determined to be $T=174 \pm 7$ MeV,
$\mu_B=46 \pm 6$ MeV, with a $\chi^2/{\rm d.o.f.} \sim 0.35$.  The parameters
are the same as those in our paper \cite{pbm2001plb}, even though a new ratio
is included and another significantly revised.  The low $\chi^2/{\rm d.o.f.}$
can be attributed to the large systematic errors in the first measurements at
RHIC; if the fit is performed with only statistical errors in the experimental
values, $\chi^2/{\rm d.o.f.}$ increases to $\sim 1.6$.  The true $\chi^2/{\rm
d.o.f.}$ lies between these two values.

\begin{figure}
\begin{center}
\includegraphics[width=13cm]{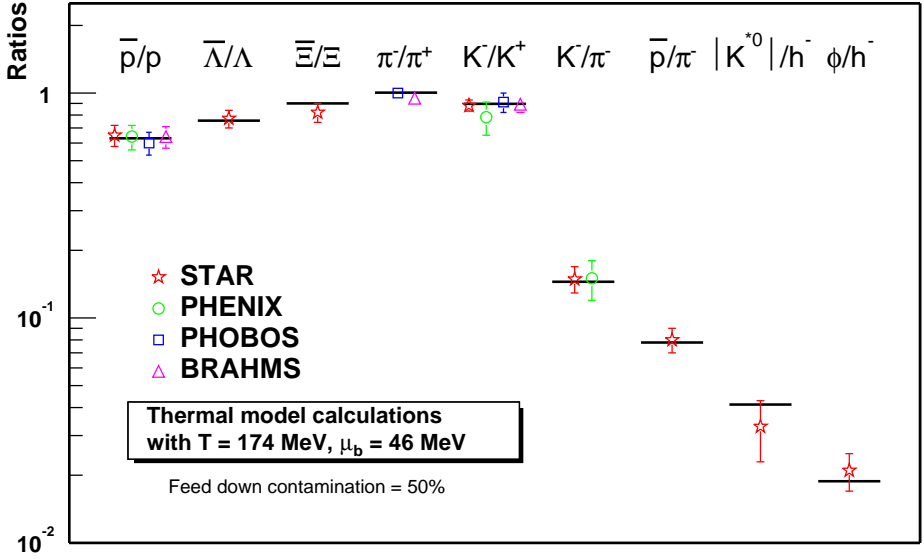}
\end{center}
\caption{Comparison between experimental data in Au+Au collisions at
$\sqrt{s}=130$ GeV and statistical model calculations of particle ratios. The
parameters of the model are $T=174$ Mev, $\mu_B=46$ MeV, and were determined
by finding the parameter set with minimum $\chi^2$. Experimental data are
taken from
\cite{STARphiprl,morrisonSQM,fachiniSQM,starprl1,zajc2001qm,phobosprl1,brahmsprl1,harris2001qm,caines2001qm,ohnishi2001qm}.}
\end{figure}

Figure 1 shows the model values for $T=174$ MeV, $\mu_B=46$ MeV, and the
available  data for the four experiments at RHIC.  Note that many of the
ratios are preliminary.  Model predictions for ratios involving $h^-$ in the
denominator have increased slightly due to the corrected treatment of negative
hadrons.  Now, as in the experimental values, only the primary negative
hadrons ($\overline{p}$, $\pi^{-}$ and $K^{-}$) are included.  In addition to
the new and updated ratios mentioned in section \ref{sec:expt}, experimental
values are taken from
\cite{starprl1,zajc2001qm,phobosprl1,brahmsprl1,harris2001qm,caines2001qm,ohnishi2001qm};
see the table of compiled ratios in our paper \cite{pbm2001plb} for specific
references. For the model calculations, a feeddown correction factor of 0.5
was applied for weak decays; see section \ref{subsec:feeddown} for a
discussion of this factor.

Two observations can be made regarding Figure 1.  First, the experimental value
for $|K^{\star0}|/h^{-}$ now lies under the model value but within error
bars.  The previous underprediction of $|K^{\star0}|/h^{-}$ was a hot topic of
discussion during this conference.  Second, the good agreement between model
and experiment for $\phi/h^{-}$ is a success of the statistical model; the
model value can be treated as a prediction, since the model parameters here
have not changed from our paper \cite{pbm2001plb}.

It should also be noted that preliminary values were shown \cite{VanBurenSQM}
at this conference by the STAR collaboration for $\Lambda/\pi^{-}$ and
$\Xi^-/\pi^{-}$.  The $\Lambda/\pi^{-}$ measurement agrees well with the
prediction from present model calculations ($\Lambda/\pi^{-}$ : expt $= 0.066
\pm 0.004_{\rm{stat}} \pm 0.005_{\rm{sys}}$; model $= 0.059$).  However,
$\Xi^-/\pi^{-}$ is underpredicted by a factor of 2 ($\Xi^-/\pi^{-}$ : expt $=
0.014 \pm 0.001_{\rm{stat}} \pm 0.003_{\rm{sys}}$; model $= 0.072$).  If the
experimental value for $\Xi^-/\pi^{-}$ holds, the large disagreement with the
statistical model would be an exciting result.  However, we stress that the
experimental results were presented as preliminary.

\subsection{Influence of feeddown correction}\label{subsec:feeddown}

As discussed in \cite{pbm1999plb} primary particle multiplicities are affected
by contributions from weak decays.  In order to compare model calculations of
particle production with experimental data, the ``contamination" from these
weak decays needs to be accounted for.  This is done in the model by
multiplying the branching ratios for weakly decaying particles ($\Lambda$,
$\Xi$, $\Omega$, etc.) by a correction factor.

Correction factors can be applied to particles individually in the model.
However, in lieu of experimental values for these quantities, the default
value used here and in our paper \cite{pbm2001plb} is 0.5, which means {\it
e.g.} that half of the protons arising from $\Lambda$ decays are not removed
from the proton multiplicity.

\begin{figure}
\begin{center}
\includegraphics[width=9cm]{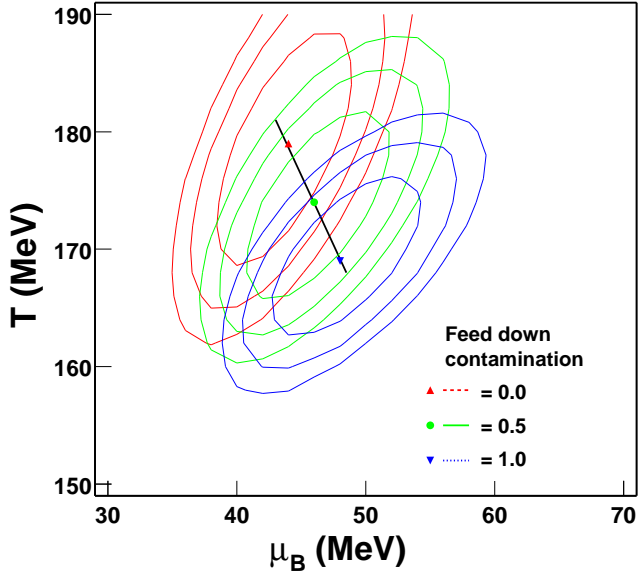}
\end{center}
\caption{Minimum $\chi^2$ points for three values of the feeddown
contamination factor.  Surrounding each minimum are lines of constant
$\chi^2$.  A contamination factor of 1 indicates that no feeddown
contributions have been removed from the ``fed" particle multiplicities.}
\end{figure}

Figure 2 shows the location of minimum $\chi^2$ on the $T-\mu_B$ diagram for
three values of the weak decay correction factor.  The lines surrounding each
minimum are contours of constant $\chi^2$.  The temperature decreases as the
contamination from weak decays increases, while $\mu_B$ changes very little.
The decrease in $T$ can be explained by the direct relation between
temperature and particle production in the statistical model.  As the
multiplicity of primary particles ($p$, $\overline{p}$, $n$, $\pi^{\pm}$, etc.) is
increasingly contaminated, fewer {\it true} protons and pions need to be
produced in order to agree with experimental values.

Of course, it is not realistic that all feeddown correction factors have the
same value.  The result of Figure 2 merely demonstrates that weak decay
contamination has a nontrivial effect on the extracted model parameters,
particularly the temperature.

\subsection{Ratios which constrain model parameters}

In this section we investigate the contribution of individual ratios to the
good agreement between the statistical model and experimental data.  The
question arises as to which ratios are strongly dependent on the model
parameters ($T$, $\mu_B$).  A ratio's sensitivity to these parameters comes
from the mass and quantum numbers of the two particle species.  This can be
expressed in the Boltzmann approximation as:
\begin{equation}\label{eq1}
{n_1\over n_2} \sim {g_1\over g_2} \Big({m_1 \over m_2}\Big)^{3/2}\
{{e^{(\mu_1-m_1)/T} }\over{e^{(\mu_2-m_2)/T} }} ,\ \ \mu_i = \mu_B B_i - \mu_S
S_i - \mu_I I_i,
\end{equation}
where $g_i$ is the spin-isospin degeneracy factor ($g_i = (2S_i+1)(2I_i+1)$),
and $B$, $S$, and $I$ represent the baryon, strangeness and isospin quantum
numbers, respectively.  The chemical potential $\mu_i$ is the sum of the
individual chemical potentials in the system.

\begin{figure}[p]
\begin{center}
\includegraphics[width=12cm]{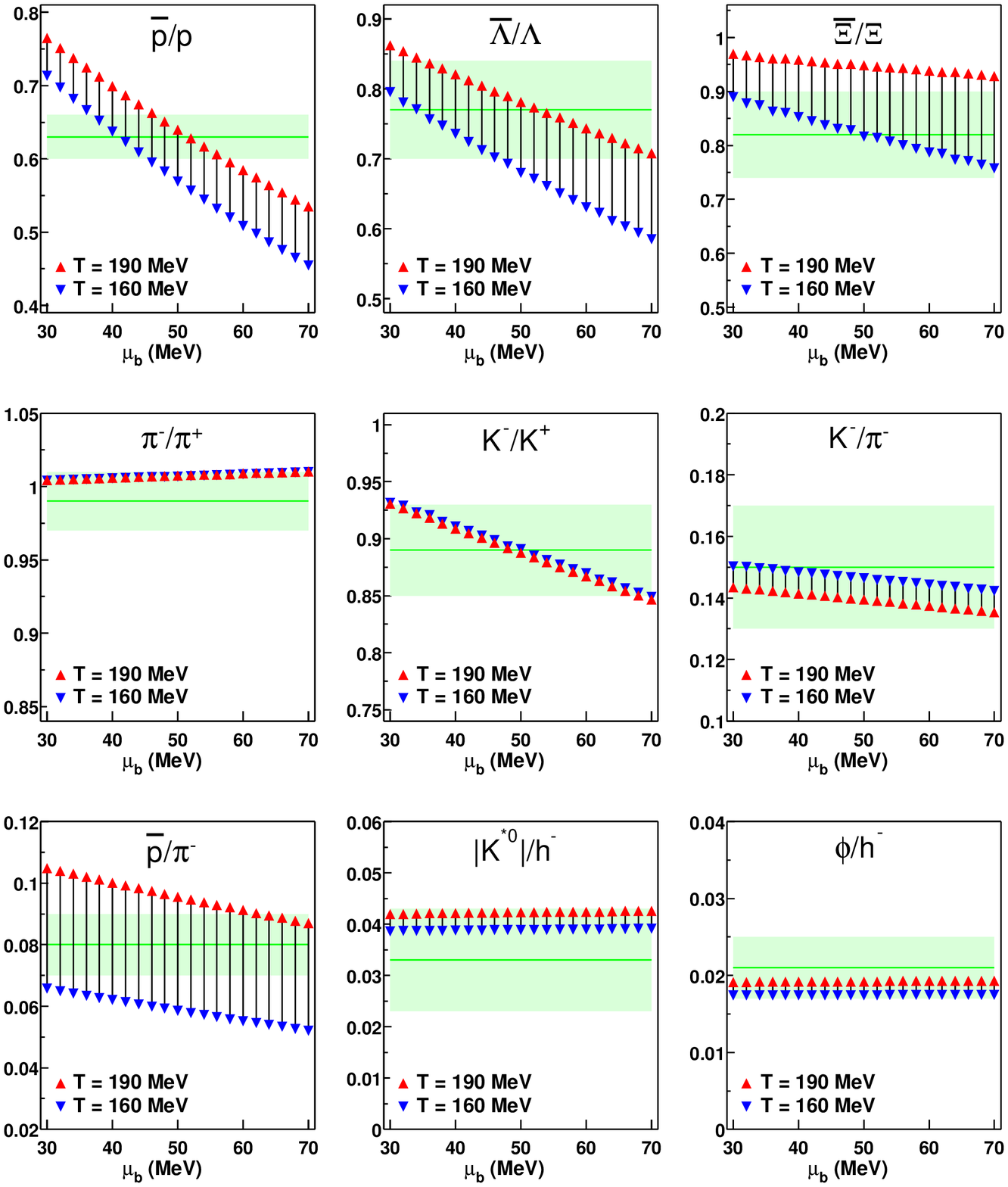}
\end{center}
\caption{Calculations with the statistical model showing the dependence on
$\mu_B$ for each ratio in Figure 1 for two different values of the
temperature.  Experimental values, compiled when multiple measurements are
available, are represented by horizontal lines; shaded areas indicate the
experimental error.}
\end{figure}

Figure 3 shows the model calculation for each ratio included in Figure 1 as a
function of $\mu_B$ for two values of the temperature.  The range of values is
chosen to include the fitted parameter values at $\sqrt{s}=130$ GeV.  The
$\overline{p}/p$, $\overline{\Lambda}/\Lambda$, and $K^{-}/K^{+}$ ratios are
strongly dependent on $\mu_B$.  In the case of $\overline{p}/p$, equation
\ref{eq1} becomes
\begin{equation}\label{eq2}
{n_{\overline{p}}\over n_p} \sim e^{-2\mu_B/T}
\end{equation}
The $\mu_B$ dependence of anti-baryon/baryon ratios is strong whenever $\mu$ is
much less than $T$; therefore, the dependence weakens for strange and
multistrange baryons due to the additional $\mu_S$ term in the chemical
potential.

\begin{figure}[p]
\begin{center}
\includegraphics[width=12cm]{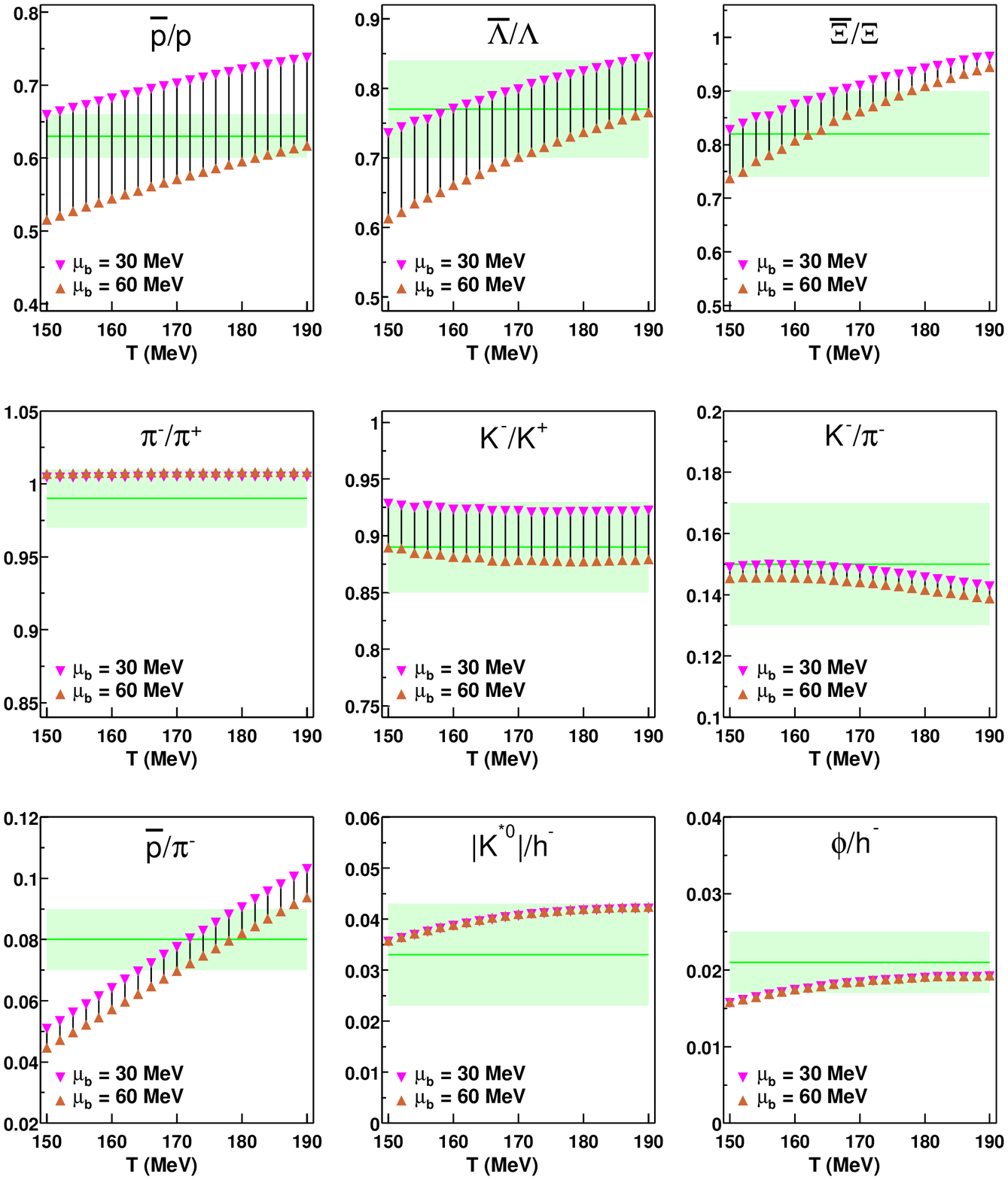}
\end{center}
\caption{Calculations with the statistical model showing the temperature
dependence of each ratio in Figure 1 for two different values of $\mu_B$.
Experimental values, compiled when multiple measurements are available, are
represented by horizontal lines; shaded areas indicate the experimental error.}
\end{figure}

Conversely, figure 4 shows the temperature dependence of each ratio for two
values of $\mu_B$. The  $\overline{p}/\pi^{-}$ ratio exhibits the strongest
dependence on the temperature.  The strong $T$ dependence of
$\overline{p}/\pi^{-}$ is due to the large mass difference between the two
species, which dominates for small values of $\mu$.  This is also the primary
reason why multi-strange baryon/meson ratios are sensitive to the temperature,
which we demonstrated in our paper \cite{pbm2001plb}.

The $\overline{\Xi^+}/\Xi^-$ ratio's dependence on temperature results from the
large contribution of the $\mu_S$ term in the chemical potential
($S_{\Xi^-}=-2$). Equation \ref{eq1} is expressed as:
\begin{equation}
{n_{\overline{\Xi^+}}\over n_{\Xi^-}}  \sim \exp \bigg( {{-2 \mu_B - 4
\mu_S}\over{T}} \bigg)
\end{equation}
Note that strangeness conservation in the collision does not require that
$\mu_S$ is zero.

\section{Conclusions}

In this contribution, the work of our previous paper \cite{pbm2001plb} was
extended to include new and updated ratios from the RHIC experiments.  The
best-fit parameters of the statistical model calculations do not change when
including the new data.  The strong agreement between model and data for the
available ratios supports the notion of a system in chemical equilibrium at
freeze-out.

Also, the uncorrected feeddown contamination from weak decays has a small but
relevant effect on the values of the extracted parameters.  The model
temperature decreases as feeddown contamination of primary particles ($p$,
$\pi$, etc.) increases because less primaries need to be produced by the model
to account for the experimental value.  Finally, $\mu_B$ was shown to be
sensitive to anti-baryon/baryon ratios, primarily $\overline{p}/p$, and the
$\overline{p}/\pi^-$ ratio shows a strong temperature dependence.

This work was done in collaboration with P. Braun-Munzinger, K. Redlich, and
J. Stachel.  The author thanks H. Van Hees and J. Knoll for fruitful
discussions regarding the statistical equations, and P. Braun-Munzinger for his
careful reading of the manuscript.

\end{document}